# Immunological recognition by artificial neural networks

**Jin Xu**

***Asia Pacific Center for Theoretical Physics, Pohang 37673***

***Department of Physics, Pohang University of Science and Technology, Pohang 37673***

**Junghyo Jo***

***School of Computational Sciences, Korea Institute for Advance Study, Seoul 02455***

***Department of Statistics, Keimyung University, Daegu 42601***

The binding affinity between the T-cell receptors (TCRs) and antigenic peptides mainly determines immunological recognition. It is not a trivial task that T cells identify the digital sequences of peptide amino acids by simply relying on the integrated binding affinity between TCRs and antigenic peptides. To address this problem, we examine whether the affinity-based discrimination of peptide sequences is learnable and generalizable by artificial neural networks (ANNs) that process the digital experimental amino acid sequence information of receptors and peptides. A pair of TCR and peptide sequences correspond to the input for ANNs, while the success or failure of the immunological recognition correspond to the output. The output is obtained by both theoretical model and experimental data. In either case, we confirmed that ANNs could learn the immunological recognition. We also found that a homogenized encoding of amino acid sequence was more effective for the supervised learning task.

PACS number: 87.15.Aa, 87.18.−h, 87.18.Sn

Keywords: T-cell receptor diversity, immunological recognition, artificial neural networks

Email: jojunghyo@kmu.ac.kr

## I. INTRODUCTION

Adaptive immunity (AI) and artificial intelligence (AI) are analogous learning processes. The immune system generates distinct T cells that express unique receptors through stochastic gene recombination, and select part of them in the thymus to avoid recognizing self-molecules [1]. Once the selected T cells are activated by certain antigenic molecules, the immune system memorizes the experience by keeping the specific T cells as memory cells to quickly recognize the experienced antigenic molecules in the future. The evolution and adaptation of immunological receptors have indeed inspired the development of a computational algorithm for pattern recognition [2, 3]. In general, machine learning optimizes artificial neural networks (ANNs) to successfully classify samples in a training set with the hope that the optimized feature extraction for the training set also works for classifying unseen samples in a test set. The learnability and generalizability of the supervised learning are also critical issues in adaptive immunity that are important in optimizing an effective repertoire of immune cells for discriminating seen and unseen antigenic molecules.

Despite the close analogy between the two AIs, a fundamental difference exists. Machine learning reads digital information of samples as the input. However, the immune system cannot directly read the sequence information of antigenic molecules, and instead mainly relies on the coarse-grained information of overall binding affinity between immunological receptors and antigenic molecules. The binding energy based model cannot directly scan the information of each amino acid site. How can T cells identify the digital sequences of peptide amino acids by simply replying on the integrated binding affinity between T-cell receptors (TCRs) and antigenic peptides? With this question in mind, we examine whether the affinity-based discrimination of peptide sequences can be learnable and generalizable by ANNs.

In addition to the recent revolution of machine learning, high-throughput sequencing technologies are able to provide sequence data from TCRs and antigenic peptides [4-7]. However, it is still challenging to experimentally obtain large-scale data for the immunological recognition between TCRs and peptides on major histocompatibility complex (pMHC), although a few hundred TCR-pMHC complex structures are available in the Protein Data Bank [8]. T-cell reactivity prediction is important for vaccine design [9-11], which is also related to a generalized problem in molecular biology about ligand-receptor interaction and even protein-protein interaction. Some pioneer work applies machine learning methods to predict T-cell reactivity by considering part of the physicochemical properties [12-14]. In this study, we adopt a simple string model [15] to determine the T-cell reactivity based on the pairwise binding affinity between TCRs and peptides, because this minimal model is sufficient to realize the integrated binding affinity. The string model has been used to explain diverse immune questions including self-nonself discrimination [16] and alloreactivity [17]. In addition to the theoretical model, we also use experimental data [5, 18, 19] to define the reactive pairs of TCRs and peptides. Given the pairs of TCR and peptide sequences as input and their reactivity as output, we examine whether ANNs can process the digital sequence information and learn the affinity-based reactivity.

Amino acids can be classified as strong (moderate) and weak ones, based on their binding energy interacting with all other amino acids. We adopt two binary encoding schemes for amino acid sequences: one considers the relative interaction strengths of amino acids, whereas the other homogenizes the differences. Then, we found that ANNs showed a higher learnability with homogenized encoding scheme. In the differentiated encoding scheme, strongly-interacting amino acids dominantly determine the output of TCR-peptide reactivity by degenerating other input elements of weakly-interacting amino acids. Natural TCRs are mostly composed of moderately-interacting amino acids since the thymic selection excludes TCRs that have strongly-interacting amino acids [15]. This suggests that the exclusion of strongly-interacting amino acids can be an effective strategy for TCRs to scan the digital information of antigenic peptide sequences.

## II. METHODS

## 1. Multilayer neural networks

We realize the task of immunological recognition by using a feedforward neural network (Fig. 1A). The neural network has inputs of amino acid sequences of TCRs and peptides. The sequence information is propagated to a hidden layer, and then the processed information finally determines an output that represents either success or failure of the TCR-peptide binding. Fully connected ANNs do not exclude self-interactions between receptor sequence amino acids and between peptide sequence amino acids. Therefore, we consider two partially connected ANNs where each amino acid of the TCR (peptide) sequence interacts with every amino acid of the peptide (TCR) sequence, but not with amino acids of the TCR (peptide) sequence (Fig. 1B). The corresponding network structure is that, in Fig. 1B, the TCR (peptide) amino acids in green (orange) are connected to the hidden layer node one by one, while the peptide (TCR) amino acids in orange (green) are fully connected to the blue hidden nodes.

## 2. Binary encoding of amino acid sequences

We used published data of TCR and antigenic peptide amino acid sequences. The TCR sequences are naïve coding/in-frame CD4+ T cell beta chain data [4], which is abstracted from the blood samples of nine human individuals [20, 21]. The database contains $10^4$-$10^5$ sequences of TCRs per person. Although the database is obtained from only nine individuals, it has been reported that the nine individuals have similar TCR diversities [4], suggesting that this TCR dataset can represent the diversity of human T-cell repertoire usually by approximately $5\times10^6$ distinct T cells [6]. In addition, we used another published dataset of dominantly naïve coding/in-frame CD8+ T cell alpha and beta chains [7], which was obtained from six human individuals. First, we translated the nucleotide sequences of DNA into corresponding amino-acid sequences by using the DNA codon table. Complementary-determining region 3 (CDR3), a specific region of TCRs, is critical in recognizing pMHCs. It is defined from the conserved cysteine around the end of the V segment to the third amino acid to the conserved phenylalanine in the J segment by the dataset [4].

For the antigenic peptide sequences, we obtained human linear infectious/autoimmune disease peptides with major histocompatibility complex (MHC) restriction from the open source "Immune epitope database and analysis resource" (IEDB) [5]. To avoid the effect of sequence lengths and use a fixed number of input nodes in ANNs, we used about $10^4$ sequences of CDR3 per person and about $10^4$ sequences of peptides with their most frequent lengths ($L$=12 for TCR and $L$=15 for peptide). From the IEDB database, we chose "Epitope: linear Epitope" to fit with the string model which will be introduced in the following section "Preparation of output data", and selected antigens with "Antigen-Organism: Hepatitis C virus (ID 11103)" and "Mycobacterium tuberculosis (ID 1773)". Here, we focused on Human TCRs by selecting "Host: Humans, Assay: T Cell Assays" with any MHC restrictions. To compare different infectious diseases later in Fig. S1, the selection criteria is supposed to include "Disease: Infectious Disease". (The list is shown after removing some sequences with errors.) Note that the selected peptides are not constrained to have known specific TCRs in IEDB.

To put the amino acid sequences into input nodes, we adopted two binary encoding schemes. First, to have equivalent descriptions for each amino acid, we use a 20-bit "one-hot" code (homogenized encoding). Then, each amino acid is represented by 20 input nodes where one node has 1, while the other nineteen nodes have 0s. The position of 1 depends on the type of amino acids. However, real amino acids have different interaction strengths. One may order 20 amino acids from weakly-interacting one to strongly-interacting one (K, D, E, …, I, F, L). Then, we can assign 5-bit binary codes (00000, 00001, 00010, …, 10001, 10010, 10011) that correspond to the sorted 20 amino acids (differential encoding). Note that strongly-interacting amino acids have more numbers of 1 that allow them to interact more with other amino acids. To assess the amino acids with binary bits, Fig. 1C schematically shows how the networks corresponds.

### 3. Preparation of output data

The success or failure of the immunological recognition corresponds to the output for the natural network. We used two methods to determine the immunological recognition: theoretical model by considering the integrated binding affinity between TCRs and peptides, and experimental data.

*A. Theoretical model*

To realize the integrated binding affinity between CDR3s of TCRs and pMHCs, we adopted a string model [15], in which the binding energy between a pair of CDR3 and pMHC sequences is defined as

$$E(\vec{t},\vec{p}) = \sum_{i=1}^{L} J(t_i, p_i) + E_c. \qquad (1)$$

The first term describes the pairwise interaction between the amino acid sequences of a CDR3 $\vec{t} = (t_1, t_2, \cdots, t_L)$ and a peptide $\vec{p} = (p_1, p_2, \cdots, p_L)$, given a length $L$ (Fig. 2A). The interaction strength between amino acids is defined by the statistical potential between each pair of amino acids, i.e., the $20 \times 20$ Miyazawa-Jernigan matrix $J$ [22, 23] (Fig. 2B). To define the binding between sequences of different lengths (12-mer TCRs and 15-mer peptides), we allowed sliding and reversing of the 12-mer TCR on the 15-mer peptide to match the sequence length of $L$=12. The second term represents the interaction between remaining peptide region/MHC and the part of TCR other than CDR3, namely CDR1 and CDR2 regions. For simplicity, $E_c$ is assumed as constant of which value can consider the differences of MHCs [24]. Here, we assumed $E_c = 33k_BT$ [15, 25], where $k_B$ is the Boltzmann constant and $T$ is the temperature. If the binding energy exceeds a threshold, $E(\vec{t},\vec{p}) > E_R$, the immunological recognition between the TCR $\vec{t}$ and the peptide $\vec{p}$ is defined as success. Here we set the threshold $E_R = 4k_BTL + E_c$ as used in the thymic selection process [15]. We confirmed that the success or failure of the immunological recognition originated from the amino acid compositions of peptides. Recognizable peptides include more strongly-interacting amino acids (Fig. 2C).

It is noteworthy that the real recognition of TCRs and peptides should be determined by their three-dimensional protein structures and dynamics. In fact, many factors such as the surface shapes of TCRs [26], loops of CDRs [27], and a few hot spots of amino acid resides [28-31] contribute to determine the real recognition. Nevertheless, given the short sequence length of CDR3, the linear pairwise model can be a reasonable approximation for the interaction between TCRs and peptides [15, 25], which has been used to explain various immunological problems including thymic selection [32-34]. More importantly, since our primary goal is not an accurate prediction of specific TCR-peptide recognition, the minimal model is sufficient to realize the nature of integrated binding affinity of TCR-peptide pairs.

### *B. Experimental data*

In particular, we filtered the peptides that have the most frequent sequence length ($L$=9) with known TCR specificities in the databases [5, 18, 19]. In addition to IEDB [5], we used two other databases, McPAS-TCR [19] and VDJdb [18]. We selected $L$=9 human peptides in McPAS-TCR [19] which includes infectious, cancer and autoimmune peptides. We also selected $L$=9 human peptides from VDJdb [18]. Those databases include the sequence information of the peptide-specific TCRs for either CDR3 alpha or beta chains. From the three databases, we obtained several thousand reactive TCR-peptide pairs as the experimental data. On the other hand, data of non-reactive TCR-peptide pairs is not available, since negative experimental results are usually not reported. The nonrecognizable samples only for training set are generated by theoretical model with the peptide sequences from IEDB [5]. We searched and selected the $L$=9 human peptide data with "Antigen-Organism: Hepatitis C virus (ID 11103)" and "Influenza Virus (ID 10002045)". The selection criteria is supposed to include "Disease: Any Disease".

## 4. Learning process of neural networks

As we described above, we used 10,000 pairs of TCR and peptide sequences as inputs, and success or failure of the corresponding TCR-peptide recognition as outputs. One set of input and output is called one sample. In particular, we prepared 5,000 recognizable sample data and 5,000 non-recognizable sample data. Later, we will discuss the effects of the ratio between recognizable versus non-recognizable samples. To put the sequence data into the input layer, we needed 540 nodes to encode 12-mer TCR plus 15-mer peptide amino acid sequences for the homogenized encoding (20 bits per amino acid), and 135 nodes for the differentiated encoding (5 bits per amino acid). The input data was processed into a single hidden layer with 300 nodes (fully connected network in Fig. 1A), 12 nodes (partially connected network B1 in Fig. 1), and 15 nodes (partially connected network B2 in Fig. 1). We also used a double hidden layer where the second layer has 300 nodes in addition to the first layer with 12 or 15 nodes, respectively. Finally, the processed data was used to determine the immunological recognition in the output layer that has two nodes. The target activity of the output nodes is defined as "10" for failure and "01" for the success of recognition. Given the network structure, we trained the network parameters by using the error backpropagation algorithm as a supervised learning [35]. In particular, we performed online learning by introducing samples from the 1st to the 10,000th. Here we define one round of learning with total training sample as one epoch. The test accuracy of the learning was measured after 100 epochs. To obtain the uncertainty of the test accuracy, we considered five ensembles with different 10,000 samples. We used a learning rate of 0.01 for the network with the single hidden layer, and a smaller learning rate of 0.001 for the network with the double hidden layers to avoid overfitting. For the control simulation with randomly determined outputs, we used 0.001 and 0.0001 for the single and double hidden layers networks, respectively.

We confirmed that 10,000 sample data size was sufficient for the supervised learning. A smaller sample data size of 2,000 or a larger size of 30,000 showed similar test accuracies (Fig. S2), although larger data sizes speeded up the learning processes in terms of the epochs. This was most significant with the case of two hidden layers. We also confirmed that the 50% composition of recognizable and non-recognizable samples was a good choice (Fig. S3). The control simulation with randomly determined outputs was significantly affected by the composition. When 20% recognizable TCR-peptide samples were randomly assigned, the test accuracy was about 80%, which was too high to judge whether ANNs are learnable or not. It was also difficult to observe the result differences among the diverse designed network structures and between two types of amino acid encoding schemes (homogenized encoding and differentiated encoding). Note that when we used experimental data of reactive TCR-peptide samples, we had data only for recognizable samples. Then, we adopted non-recognizable TCR-peptide samples from the theoretical model as part of the training set. Finally, once the sample data was prepared, we split them into a training set (2/3) and a test set (1/3) for learning and testing the neural networks respectively.

## III. RESULTS

### 1. Neural networks can learn affinity-based discrimination of peptide sequences

T cells discriminate peptides based the binding affinity between TCRs and pMHCs. To examine how the coarse-grained information of the integrated affinity can discriminate the digital information of peptide sequences, we study the discrimination of peptide sequences by artificial neural networks. We use multi-layer neural networks to read inputs of a pair of TCR and peptide sequences, and to predict outputs of success or failure for the recognition (i.e., strong binding) of the TCR-peptide pair. The inputs of TCR and peptide sequences are encoded by two types of binary codes. The homogenized encoding ignores the differences of the relative interaction strengths between amino acids, whereas the differentiated encoding considers the differences (see Methods for details). Here we used open databases to obtain the amino acid sequences of human TCRs and peptides for human infectious diseases (see Methods for details). The outputs for the immunological recognition were prepared using both theoretical model and experimental data. First, we present our results based on the theoretical model that predicts the success or failure of the peptide recognitions by using the integrated binding affinity between TCR and peptide sequences.

The multi-layer neural networks with a single hidden layer could learn the affinity-based discrimination of peptide sequences (Fig. 3). The fully connected network between input and hidden layer has 300 hidden nodes, but includes self-interactions between amino acid sites within a TCR sequence and within a peptide sequence. To exclude the self-interactions, we designed two types of partially connected networks, with 12 (15) hidden nodes for B1 (B2) in Fig. 1. The input nodes from the TCR sequence are fully connected with the input nodes from the peptide sequences through the hidden layer, but the input nodes from the peptides are not connected with each other. The partially connected networks showed lower test accuracy than the fully connected networks regardless of the encoding schemes for inputs (Fig. 3). To improve this, we tested another partially connected ANN design with more hidden layer nodes. For homogenized encoding, the TCR (peptide) binary codes instead of amino acids are connected to the hidden layer node one by one, while the peptide (TCR) binary codes are fully connected to the hidden layer nodes. Therefore, the number of hidden layer nodes for homogenized encoding is 240 (300) for B1 (B2) in Fig. 1. For differentiated encoding, there are 60 (75) nodes in the hidden layer for B1 (B2) in Fig. 1. When we introduced more hidden nodes in the partially connected network, their accuracies were improved, but the sparse networks still showed lower accuracy than the fully connected network. Adding a second hidden layer with 300 nodes to the first layer with 12 or 15 nodes did not improve the test accuracy, but slowed down the learning processes (Fig. 3). The learnability of neural networks represents that the affinity-based peptide discrimination has general rules applying not only for training data but also for test data. Once we randomly assigned success or failure of the recognition samples, the test accuracy was 50% as expected (Fig. 3).

We are motivated to test data from different diseases and check whether our methods are valid for use in various antigens/diseases. In addition to the virus infectious disease (hepatitis C virus disease) and the bacterial infectious disease (Mycobacterium tuberculosis diseases), neural networks could also discriminate peptides derived from the autoimmune disease (Fig. S1). The peptide sequences were obtained by searching the IEDB for specific organisms and diseases following the way described in the section of Methods. We selected "Disease: Infectious Disease/Autoimmune Disease" for infectious and autoimmune diseases. Since the immunological recognition was determined by the integrated binding affinity between amino acid sequences of TCRs and peptides, the origin of peptides did not matter, but their amino acid compositions did matter. This is why cross-reactivity exists when a self-peptide is homologous to a pathogen peptide. Furthermore, we also confirmed the learnability of affinity-based peptide discrimination by neural networks using another set of T-cell repertoire pools with both CDR3 alpha and beta chains (Fig. S4).

**2. Homogenized amino acid strengths are effective for T-cell receptors to discriminate peptides**

We considered two encoding schemes for amino acid sequences. The homogenized encoding ignores the relative interaction strengths between amino acids, whereas the differentiate encoding considers the different interaction strengths. The results by homogenized encoding showed a higher test accuracy than the differentiated encoding (Fig. 3A and Fig. 3B; Fig. S1A and Fig. S1B), although both cases are learnable. Furthermore, the homogenized encoding highlighted the differences during the training processes (Fig. 3C and Fig. 3D; Fig. S1C and Fig. S1D). By homogenized encoding, every amino acid becomes equally important and contributes to the integral binding affinity. However, by differentiated encoding, strongly-interacting amino acids can dominantly determine the integrated binding affinity, and degenerate the contributions from weakly-interacting amino acids. Therefore, the differentiated encoding has difficulties to discriminate the sequence differences. The immune system may use homogenized amino acids that minimize their different interaction strengths. During the thymic selection, the immune system excludes naïve T cells of which receptors are composed with strongly-interacting amino acids [15]. This has been proposed as “bar-code model” that explains the specificity of TCRs [15].

**3. Neural networks can learn real reactivity between T-cell receptors and antigenic peptides**

We demonstrated that neural networks could learn the affinity-based peptide discrimination defined by theoretical model. Now we examine if neural networks can also learn real reactivity between TCRs and peptides obtained from experimental data. We used experimentally known several thousand reactive pairs of TCRs and peptides in three open databases, i.e., IEDB [5], McPAS-TCR [19], and VDJdb [18] (see Methods for details). For the supervised learning, we also needed to prepare non-reactive pairs of TCRs and peptides for the training set. However, since the nonreacognizable data is not available in the experimental literatures, we adopted the non-reactive TCR-peptide samples in the previous theoretical model. To consider different amino acid strengths in reality, we only used the differentiated encoding scheme here. And to obtain high test accuracies, we used the fully connected network and partially connected networks with more hidden nodes. There are 300 nodes for network Fig. 1A and 60 (45) nodes in the hidden layer for B1 (B2) in Fig 1. Then, we confirmed that the neural networks could learn the real reactivity between TCRs and peptides (Fig. 4A and Fig. 4C). Note that the test accuracy measures only recognizable TCR-peptide reactivity, unlike the previous test accuracy in Fig. 3 and also Fig. S1-Fig. S4 that consider both recognizable and nonrecognizable samples.

Finally, we examined if the neural networks, trained by the previous theoretical model, can correctly predict the experimentally observed TCR-peptide reactivity. The test accuracies for the recognizable samples were not high (<50%). Thus, unless the neural network is newly trained by the experimental data, the optimized networks for the theoretical model cannot fully explain the real TCR-peptide reactivity. In conclusion, neural networks can learn the affinity-based peptide discrimination defined either by theoretical model or by experimental data. However, the training is not transferable between the two methods. This implies that the real affinity of TCR-peptide reactivity cannot be sufficiently explained simply by the string model.

## IV. DISCUSSION

The immune system discriminates different peptides relying on the binding affinity between T-cell receptors and peptides. The integrated binding affinity should be a coarse-grained information compared with the microscopic information of amino acid sequences of peptides. Can the coarse-grained information be sufficient to represent the microscopic information? With this question in mind, we examined if neural networks could process the microscopic information, and predict the immunological recognition based on the coarse-grained information.

We used large-scale high-throughput data of TCRs [4, 7] and open-source peptide amino acid sequence data [5] as the input. The success or failure of the recognition between TCRs and peptides are set as our output. We used a theoretical model to define the integrated binding affinity between TCRs and peptide sequences [15], and also experimental data of reactive pairs of TCRs and peptides [5, 18, 19]. Given the input and output data, feedforward neural networks could learn the input-output mapping. This demonstrated that the neural networks could realize the affinity-based immunological recognition.

People have considered immunology as an information process [36] with binary strings presenting TCRs and peptides. Our study considered immune receptors and pMHCs at a more detailed scale and transformed each amino acid into binary bits. Therefore, the ANNs can catch the information of each amino acid site rather than the integral binding affinity. We adopted two binary encoding schemes for amino acid sequences. The homogenized encoding ignores the relative interaction strengths of amino acids, whereas the differentiated encoding considers the differences. Then, we showed that the homogenized encoding was more effective to learn the immunological recognition, because strongly-interacting amino acids dominated to determine the integral binding affinity by degenerating the contributions from weakly-interacting amino acids in the differentiated encoding. Naïve T cells are removed in the thymus if their TCRs have too strongly-interacting or weakly-interacting amino acids [15]. The thymic selection may contribute to the homogenized encoding for TCRs, which helps scan the microscopic information of peptide sequences.

We demonstrated that the coarse-grained information of the integrated binding affinity for peptides could sufficiently represent the microscopic information of peptide sequences by using neural networks. After putting some experimental proved immunological recognition data into the training set, our work shows potential for T-cell reactivity prediction. If there could be more immune recognizable data and especially non-recognizable data, the accuracy could be more reliable. However, our simple method only represents the binding between TCR and peptide sequences related to amino acid strengths, which cannot fully understand the real reactivity between TCRs and peptides relying on their three-dimensional structures and dynamics.

## ACKNOWLEDGEMENT

The study was supported by the Basic Science Research Program of the National Research Foundation of Korea (NRF, www.nrf.re.kr), the Ministry of Education (2016R1D1A1B03932264) to JJ. The funders had no role in study design, data collection and analysis, decision to publish, or preparation of the manuscript.

## REFERENCES

[1] N. K. Jerne, Science **229**, 1057 (1985).

[2] D. Dasgupta, Ieee Sys Man Cybern, 873 (1997).

[3] J. D. Farmer, N. H. Packard and A. S. Perelson, Physica D: Nonlinear Phenomena **22**, 187 (1986).

[4] A. Murugan, T. Mora, A. M. Walczak and C. G. Callan, Jr., Proc. Natl. Acad. Sci. U. S. A. **109**, 16161 (2012).

[5] R. Vita, J. A. Overton, J. A. Greenbaum, J. Ponomarenko, J. D. Clark et al., Nucleic Acids Res. **43** (Database issue), D405(2015).

[6] R. L. Warren, D. Freeman, T. Zeng, G. Choe, S. Munro et al., Genome Res. **21** (5), 790 (2011).

[7] I. V. Zvyagin, M. V. Pogorelyy, M. E. Ivanova, E. A. Komech, M. Shugay et al., Proc. Natl. Acad. Sci. U. S. A. **111**, 5980 (2014).

[8] H. M. Berman, J. Westbrook, Z. Feng, G. Gilliland, T. N. Bhat et al., Nucleic Acids Res. **28**, 235(2000).

[9] N. C. Toussaint, P. Donnes and O. Kohlbacher, PLoS Comput. Biol. **4**, e1000246 (2008).

[10] N. C. Toussaint and O. Kohlbacher, Nucleic Acids Res. **37** (Web Server issue), W617 (2009).

[11] T. Vider-Shalit, S. Raffaeli and Y. Louzoun, Mol. Immunol. **44**, 1253 (2007).

[12] T. Saethang, O. Hirose, I. Kimkong, V. A. Tran, X. T. Dang et al., J. Immunol. Methods **387**, 293 (2013)

[13] C. W. Tung and S. Y. Ho, Bioinformatics **23**, 942 (2007).

[14] C. W. Tung, M. Ziehm, A. Kamper, O. Kohlbacher and S. Y. Ho, BMC Bioinformatics **12**, 446 (2011).

[15] A. Kosmrlj, A. K. Jha, E. S. Huseby, M. Kardar and A. K. Chakraborty, Proc. Natl. Acad. Sci. U. S. A. **105**, 16671 (2008).

[16] J. K. Percus, O. E. Percus and A. S. Perelson, Proc. Natl. Acad. Sci. U. S. A. **90**, 1691 (1993).

[17] V. Detours, R. Mehr and A. S. Perelson, J. Immunol. **164**, 121 (2000).

[18] M. Shugay, D. V. Bagaev, I. V. Zvyagin, R. M. Vroomans, J. C. Crawfor et al., Nucleic Acids Res. **46**, D419 (2018).

[19] N. Tickotsky, T. Sagiv, J. Prilusky, E. Shifrut and N. Friedman, Bioinformatics **33**, 2924 (2017).

[20] H. S. Robins, P. V. Campregher, S. K. Srivastava, A. Wacher, C. J. Turtle et al., Blood **114**, 4099 (2009).

[21] H. S. Robins, S. K. Srivastava, P. V. Campregher, C. J. Turtle, J. Andriesen et al., Sci. Transl. Med. **2**, 47ra64 (2010).

[22] H. Li, C. Tang and N. S. Wingreen, Phys. Rev. Lett. **79**, 765 (1997).

[23] S. Miyazawa and R. L. Jernigan, J. Mol. Biol. **256**, 623 (1996).

[24] A. Kosmrlj, M. Kardar and A. K. Chakraborty, J. Stat. Phys. **149**, 203 (2012).

[25] A. Kosmrlj, A. K. Chakraborty, M. Kardar and E. I. Shakhnovich, Phys. Rev. Lett. **103** (2009).

[26] M. S. Park, S. Y. Park, K. R. Miller, E. J. Collins and H. Y. Lee, Mol. Immunol. **56**, 81 (2013).

[27] A. K. Sewell, Nature Reviews Immunology **12**, 668 (2012).

[28] H. R. Chen, A. K. Chakraborty and M. Kardar, Phys. Rev. E **97** (2018).

[29] M. E. Birnbaum, J. L. Mendoza, D. K. Sethi, S. Dong, J. Glanville et al., Cell **157**, 1073 (2014).

[30] J. J. Calis, R. J. de Boer and C. Kesmir, PLoS Comput. Biol. **8**, e1002412 (2012).

[31] B. D. Stadinski, K. Shekhar, I. Gomez-Tourino, J. Jung, K. Sasaki et al., Nat. Immunol. **17**, 946 (2016).

[32] D. L. Chao, M. P. Davenport, S. Forrest and A. S. Perelson, Eur. J. Immunol. **35**, 3452 (2005).

[33] V. Detours, R. Mehr and A. S. Perelson, J. Theor. Biol. **200**, 389 (1999).

[34] V. Detours and A. S. Perelson, Proc. Natl. Acad. Sci. U. S. A. **96**, 5153 (1999).

[35] D. E. Rumelhart, G. E. Hinton and R. J. Williams, Nature **323**, 533 (1986).

[36] L. A. Segel and I. R. Cohen, *Design principles for the immune system and other distributed autonomous system*. (Oxford University Press, New York, 2001).

Figure Captions.

Fig. 1. (Color online) Feedforward neural networks for immunological recognitions. Amino acid sequences of a T-cell receptor (TCR, length $L$=12, filled green circles) and a peptide ($L$=15, filled orange circles) correspond to an input, and the recognition (filled black circle) of the TCR-peptide pair represents an output. Total 5,000 recognizable and 5,000 nonrecognizable samples are prepared, two third of them are assigned as a training set, and one third of them are kept as a test set. The input data is processed into hidden nodes (filled blue circles), and then the transformed data is used to predict the recognition of input pairs of TCR-peptide onto output nodes. A fully connected network between input nodes and hidden nodes is considered first (A). To exclude the self-interactions between amino acids within a TCR sequence (B1), or between amino acids within a peptide sequence (B2), partially connected networks are also considered. Each amino acid is encoded by binary codes. Homogenized encoding uses a 20-bit "one-hot" code in which twenty amino acids have only one "1" site among their 20 bits (C1). However, differentiated encoding uses a 5-bit code that considers the relative interaction strengths between amino acids (C2). The 5-bit binary codes are assigned as 00000, 00001, 00010, …, 10001, 10010, 10011 from weakly interacting amino acids to strongly interacting amino acids.

Fig. 2. (Color online) Theoretical model for defining immunological recognition. (A) Pairwise interaction between amino acid sequences of a TCR and a peptide presented on major histocompatibility complex (MHC). The integrated binding affinity is determined by two parts: specific interaction (red connections) between complementary-determining region 3 (CDR3) of the TCR and the paired peptide region; non-specific interaction (black connections) between the remaining region except for CDR3 of the TCR (CDR1 and CDR2 regions) and the remaining peptide region/MHC. (B) Miyazawa-Jernigan matrix defines the statistical interaction strengths between twenty amino acids. Amino acids are sorted from the most strongly interacting one to the most weakly one according to the maximum binding energy interacting with all other amino acids. (C) Amino acid compositions of recognizable and nonrecognizable peptides by the theoretical model. Recognizable peptides have more strongly interacting amino acids. Errors are estimated by five ensembles of 10,000 samples.

Fig. 3. (Color online) Learnability of immunological recognition by neural networks. The input amino acid sequences of TCRs and peptides are encoded by binary codes, either homogenized encoding or differentiated encoding scheme, in order to ignore or consider the relative amino acids strengths. The training accuracy of theoretical predicted recognition from input pairwise TCR-peptide is plotted with learning steps (epochs) given (A) homogenized encoding and (B) differentiated encoding. Seven

networks are considered. F (filled red diamonds) represents fully connected network between input and hidden layers (300 hidden nodes). P1 (filled red circles) and P2 (empty red circles) represent partially connected networks removing self-interactions between amino acids within a TCR sequence (12 hidden nodes) and within a peptide sequence (15 hidden nodes). M1 (filled red triangles) and M2 (empty red triangles) represent partially connected networks with more hidden nodes (240/300 hidden nodes for homogenized encoding scheme, and 60/75 nodes for differentiated encoding scheme). D1 (filled red squares) and D2 (empty red squares) represent partially connected networks with double hidden layers (12/15 nodes in the first hidden layer for D1/D2, and 300 nodes in the second hidden layer). To confirm the learnability of the immunological recognition, the outputs are randomly shuffled as the control dataset shown in black compared to the original in red. Only the first 50 epochs of the learning processes are shown, as there are no significant changes from the 50th to the 100th epoch with a sample size of 10,000. The test accuracy of the immunological recognition is also plotted for the homogenized encoding scheme (C) and differentiated encoding scheme (D) after 100 epochs of learning.

Fig. 4. (Color online) Learnability of real reactivity between T-cell receptors and antigenic peptide by neural networks. (A) The training accuracy of predicted recognition from partially experimental input pairwise TCR-peptide is plotted with learning steps (epochs) given differentiated binary encoding for amino acids. Three networks are considered. F (filled red diamonds) represents fully connected network between input and hidden layers (300 hidden nodes). M1 (filled red triangles) and M2 (empty red triangles) represent partially connected networks excluding self-interactions between amino acids within a TCR sequences (60 hidden nodes for M1) and within a peptide (45 hidden nodes for M2). The recognizable samples are from experimental TCR-peptide pairs from IEDB, McPAS-TCR and VDJdb databases [5, 18, 19]. The nonrecognizable samples only for training set are generated by theoretical model with the peptide sequences from IEDB [5]. Only the first 50 epochs of the learning processes are shown, as there are no significant changes from the 50th to the 100th epoch with a sample size of 10,000. "All diseases" (red), "Hepatitis C virus disease" (green), and "Influenza disease" (blue) are considered.

(B) The test accuracy of the immunological recognition is also plotted after 100 epochs of learning. (C) The training accuracy of the immunological recognition defined by theoretical model. (D) Given the training based on the theoretical model, the test accuracy is obtained from the experimental data of the real reactivity between TCRs and antigenic peptides.

Fig. S1. (Color online) Learnability of immunological recognition for different diseases by neural networks. The input amino acid sequences of TCRs and peptides are encoded by binary codes, either homogenized encoding or differentiated encoding scheme, in order to ignore or consider the amino acids strengths. The training accuracy of theoretical predicted recognition from input pairwise TCR-peptide is plotted with learning steps (epochs) given (A) homogenized encoding and (B) differentiated encoding. Seven networks are considered. F (filled red diamonds) represents fully connected network between input and hidden layers (300 hidden nodes). P1 (filled red circles) and P2 (empty red circles) represent partially connected networks removing self-interactions between amino acids within a TCR sequence (12 hidden nodes) and within a peptide sequence (15 hidden nodes). M1 (filled red triangles) and M2 (empty red triangles) represent partially connected networks with more hidden nodes (240/300 hidden nodes for homogenized encoding scheme, and 60/75 nodes for differentiated encoding scheme). D1 (filled red squares) and D2 (empty red squares) represent partially connected networks with double hidden layers (12/15 nodes in the first hidden layer for D1/D2, and 300 nodes in the second hidden layer). Different colors represent different diseases, i.e., red represents the autoimmune disease, green represents Hepatitis C virus disease and blue represents Mycobacterium tuberculosis disease. Only the first 50 epochs of the learning processes are shown, as there are no significant changes from the 50th to the 100th epoch with a sample size of 10,000. The test accuracy of the immunological recognition is also plotted for the homogenized encoding scheme (C) and differentiated encoding scheme (D) after 100 epochs of learning.

Fig. S2. (Color online) Learnability of immunological recognition for different sample sizes by neural networks. The input amino acid sequences of TCRs and peptides are encoded by binary codes, either homogenized encoding or differentiated encoding scheme, in order to ignore or consider the amino acids

strengths. The training accuracy of theoretical predicted recognition from input pairwise TCR-peptide is plotted with learning steps (epochs) given (A) homogenized encoding and (B) differentiated encoding. Seven networks are considered. F (filled red diamonds) represents fully connected network between input and hidden layers (300 hidden nodes). P1 (filled red circles) and P2 (empty red circles) represent partially connected networks removing self-interactions between amino acids within a TCR sequence (12 hidden nodes) and within a peptide sequence (15 hidden nodes). M1 (filled red triangles) and M2 (empty red triangles) represent partially connected networks with more hidden nodes (240/300 hidden nodes for homogenized encoding scheme, and 60/75 nodes for differentiated encoding scheme). D1 (filled red squares) and D2 (empty red squares) represent partially connected networks with double hidden layers (12/15 nodes in the first hidden layer for D1/D2, and 300 nodes in the second hidden layer). Different colors represent different sample sizes, i.e., green represents a sample data size of 2,000 and blue represents a size of 30,000. Only the first 50 epochs of the learning processes are shown, as there are no significant changes from the 50th to the 100th epoch with a sample size of 30,000. The test accuracy of the immunological recognition is also plotted for the homogenized encoding scheme (C) and differentiated encoding scheme (D) after 100 epochs of learning.

Fig. S3. (Color online) Learnability of immunological recognition for different percentages of successful recognition as the output by neural networks. The input amino acid sequences of TCRs and peptides are encoded by binary codes, either homogenized encoding or differentiated encoding scheme, in order to ignore or consider the amino acids strengths. The training accuracy of theoretical predicted recognition from input pairwise TCR-peptide is plotted with learning steps (epochs) given (A) homogenized encoding and (B) differentiated encoding. Seven networks are considered. F (filled red diamonds) represents fully connected network between input and hidden layers (300 hidden nodes). P1 (filled red circles) and P2 (empty red circles) represent partially connected networks removing self-interactions between amino acids within a TCR sequence (12 hidden nodes) and within a peptide sequence (15 hidden nodes). M1 (filled red triangles) and M2 (empty red triangles) represent partially

connected networks with more hidden nodes (240/300 hidden nodes for homogenized encoding scheme, and 60/75 nodes for differentiated encoding scheme). D1 (filled red squares) and D2 (empty red squares) represent partially connected networks with double hidden layers (12/15 nodes in the first hidden layer for D1/D2, and 300 nodes in the second hidden layer). Different colors represent different percentages of successful recognition as the output, i.e., green represents infectious diseases with 20% recognizable samples as output, and blue represents random input-output mapping with 20% recognizable samples as a control. Only the first 50 epochs of the learning processes are shown, as there are no significant changes from the 50th to the 100th epoch with a sample size of 10,000. The test accuracy of the immunological recognition is also plotted for the homogenized encoding scheme (C) and differentiated encoding scheme (D) after 100 epochs of learning.

Fig. S4. (Color online) Learnability of immunological recognition for different T-cell repertoire by neural networks. The input amino acid sequences of TCRs and peptides are encoded by binary codes, either homogenized encoding or differentiated encoding scheme, in order to ignore or consider the amino acids strengths. The training accuracy of theoretical predicted recognition from input pairwise TCR-peptide is plotted with learning steps (epochs) given (A) homogenized encoding and (B) differentiated encoding. Seven networks are considered. F (filled red diamonds) represents fully connected network between input and hidden layers (300 hidden nodes). P1 (filled red circles) and P2 (empty red circles) represent partially connected networks removing self-interactions between amino acids within a TCR sequence (12 hidden nodes) and within a peptide sequence (15 hidden nodes). M1 (filled red triangles) and M2 (empty red triangles) represent partially connected networks with more hidden nodes (240/300 hidden nodes for homogenized encoding scheme, and 60/75 nodes for differentiated encoding scheme). D1 (filled red squares) and D2 (empty red squares) represent partially connected networks with double hidden layers (12/15 nodes in the first hidden layer for D1/D2, and 300 nodes in the second hidden layer). Different colors represent different T-cell repertoire, i.e. green represents CD8+ T cell alpha chain data and blue

represents CD8+ T cell beta chain data. Only the first 50 epochs of the learning processes are shown, as there are no significant changes from the 50th to the 100th epoch with a sample size of 10,000. The test accuracy of the immunological recognition is also plotted for the homogenized encoding scheme (C) and differentiated encoding scheme (D) after 100 epochs of learning.

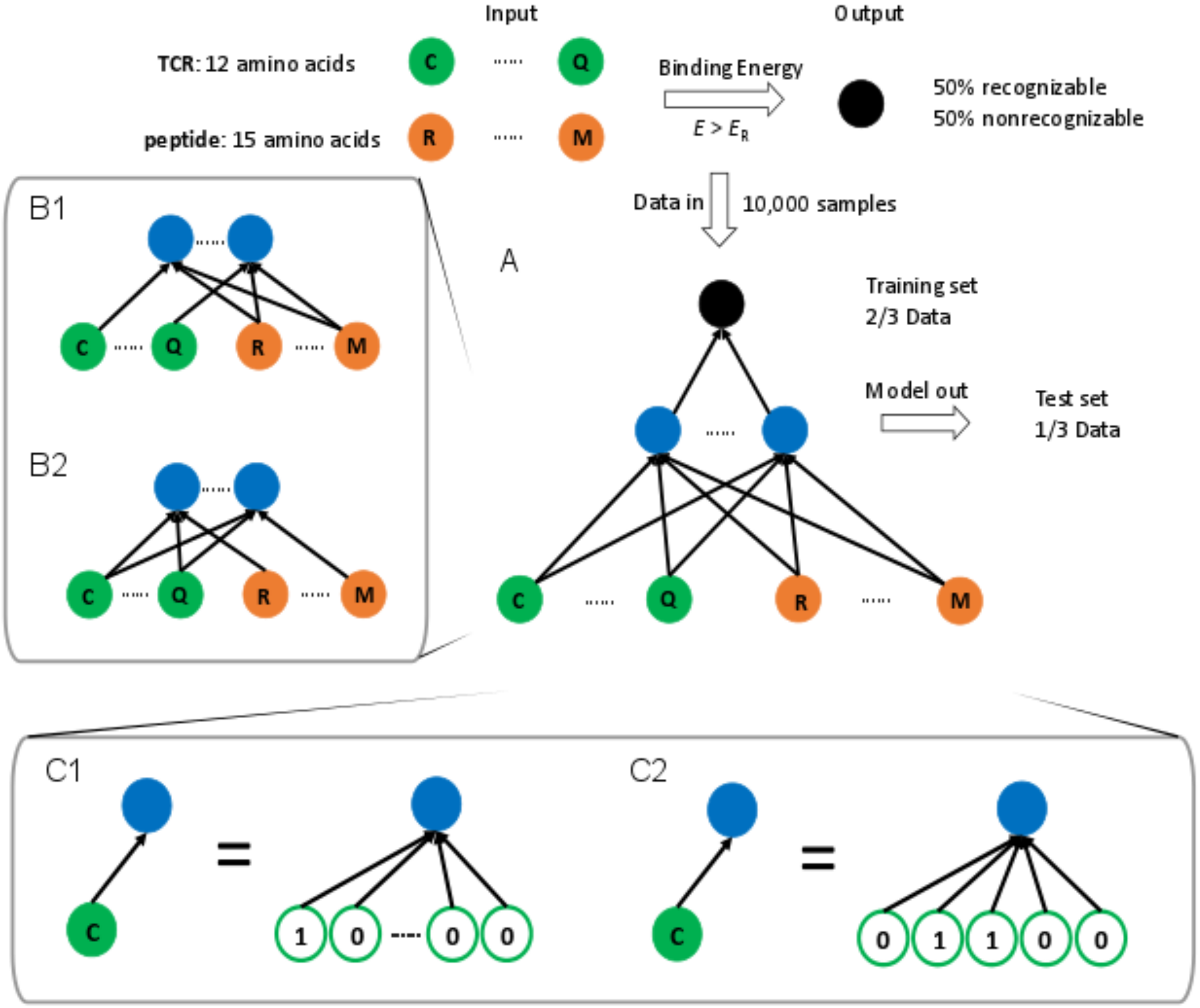

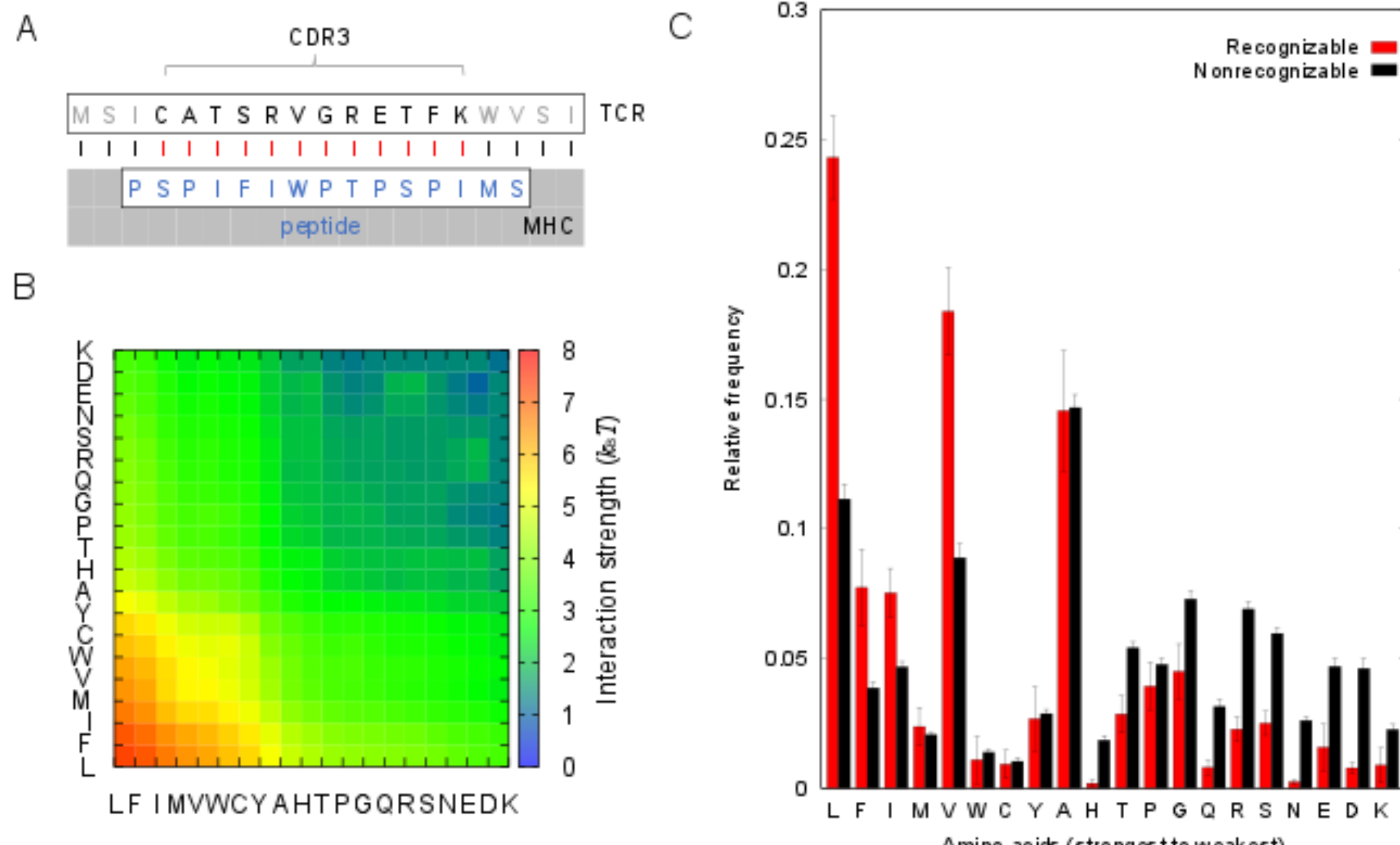

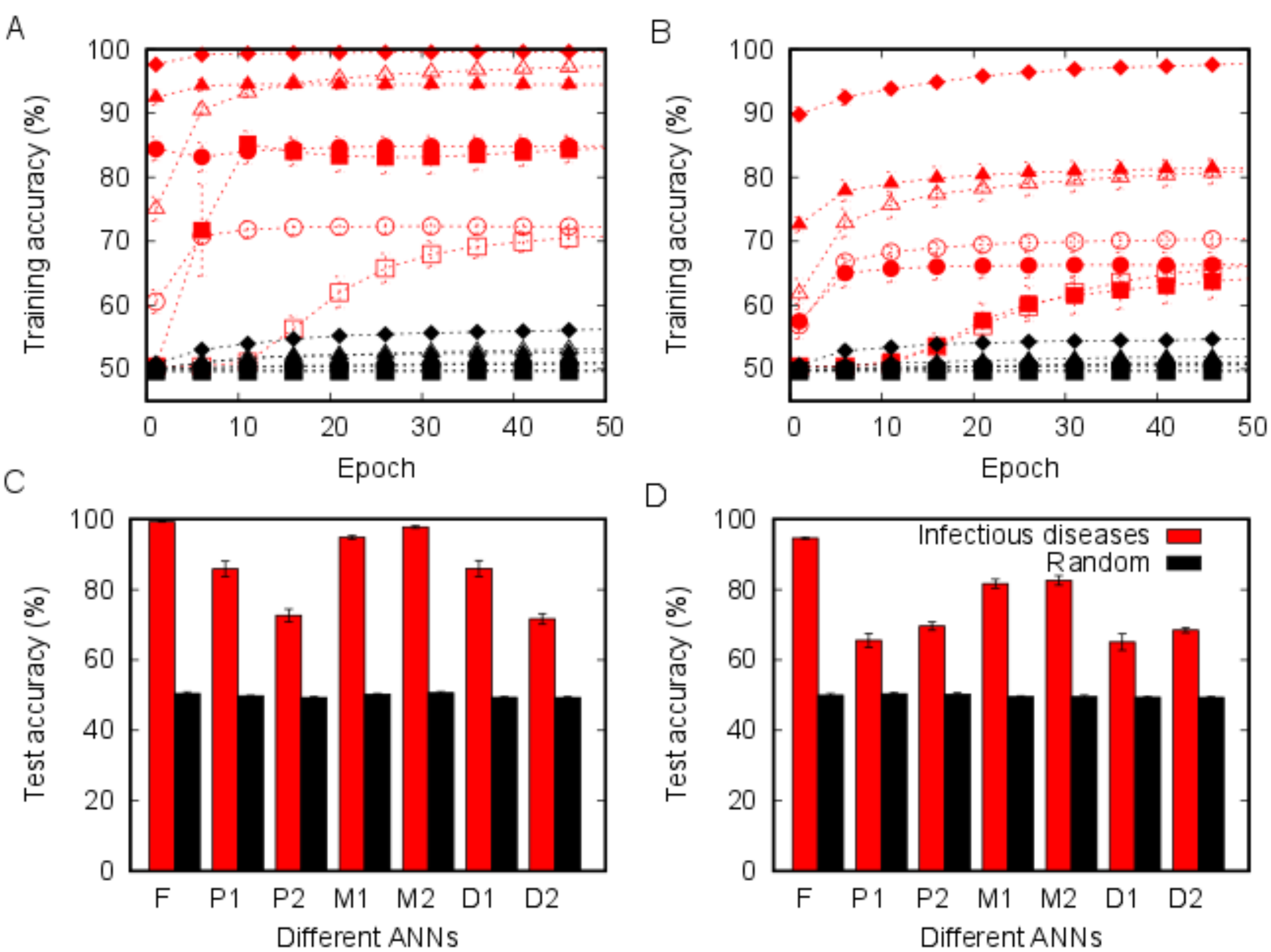

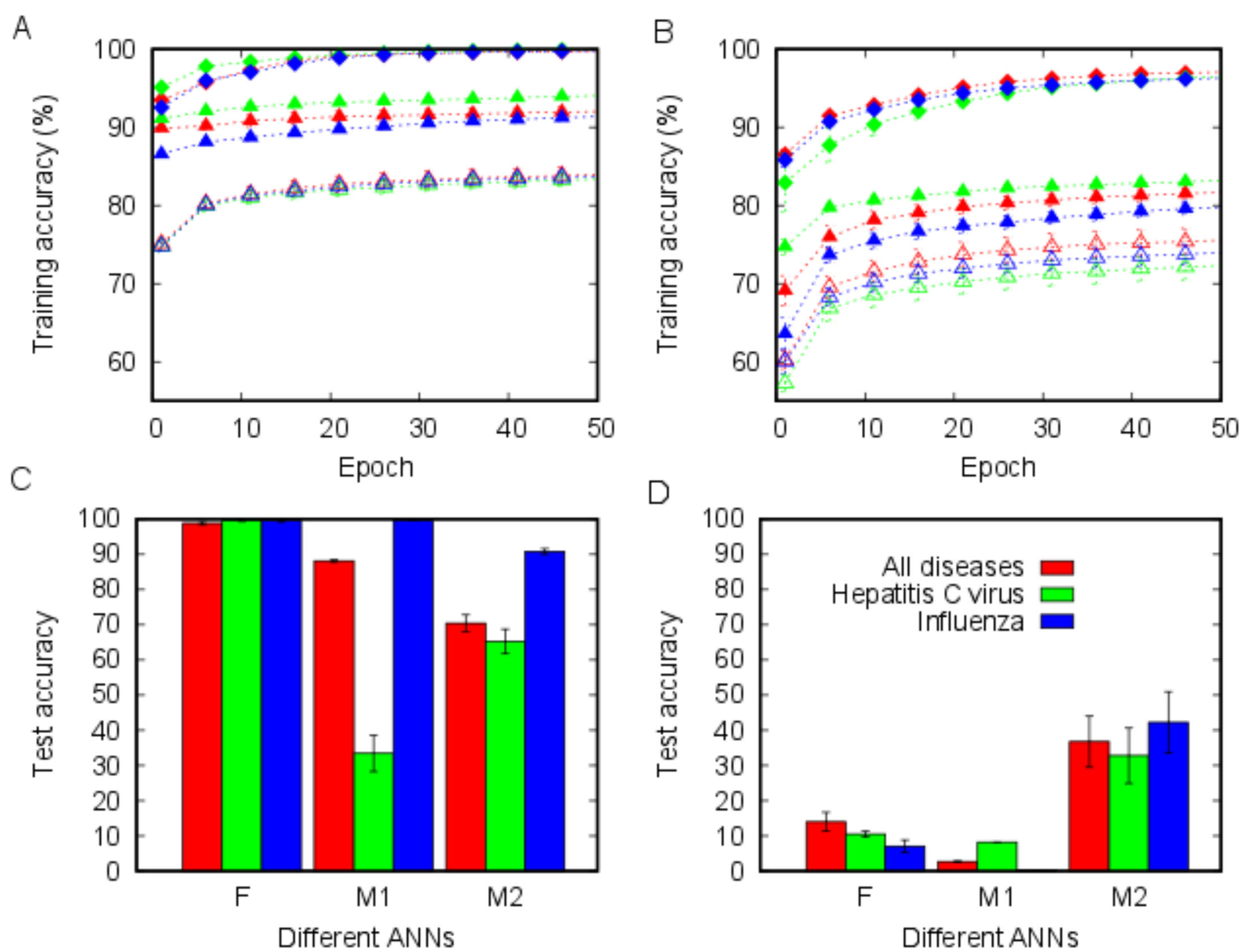

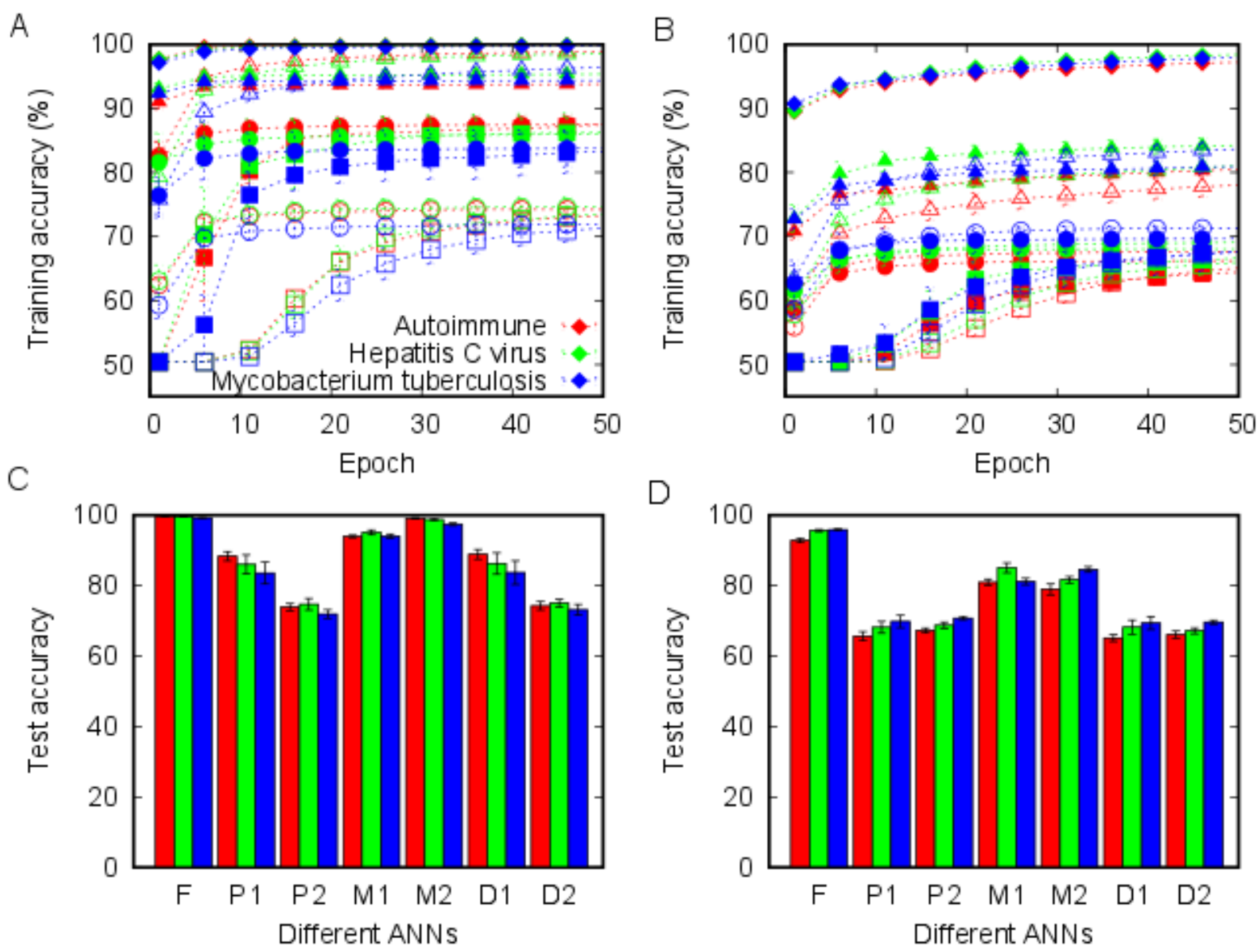

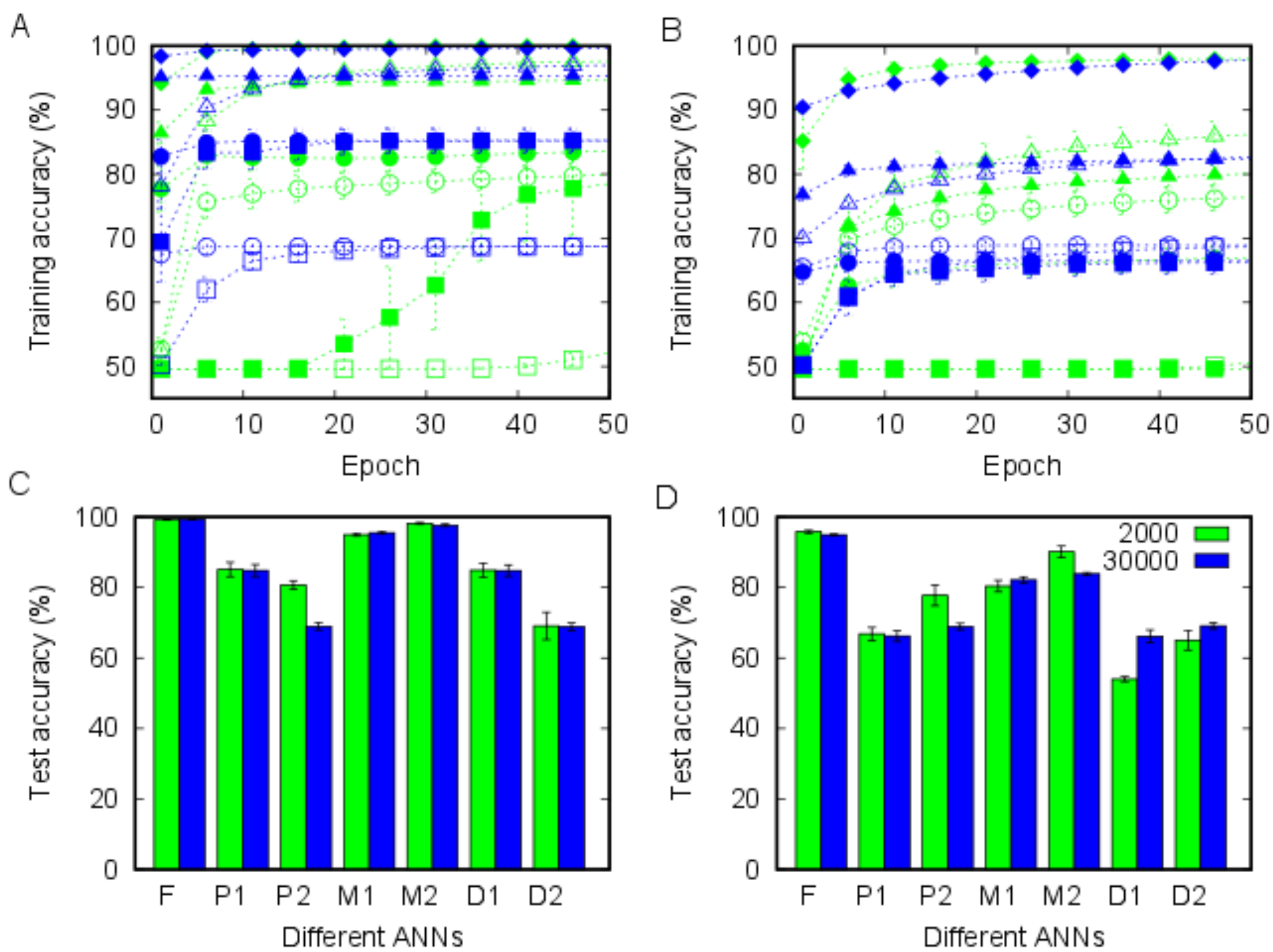

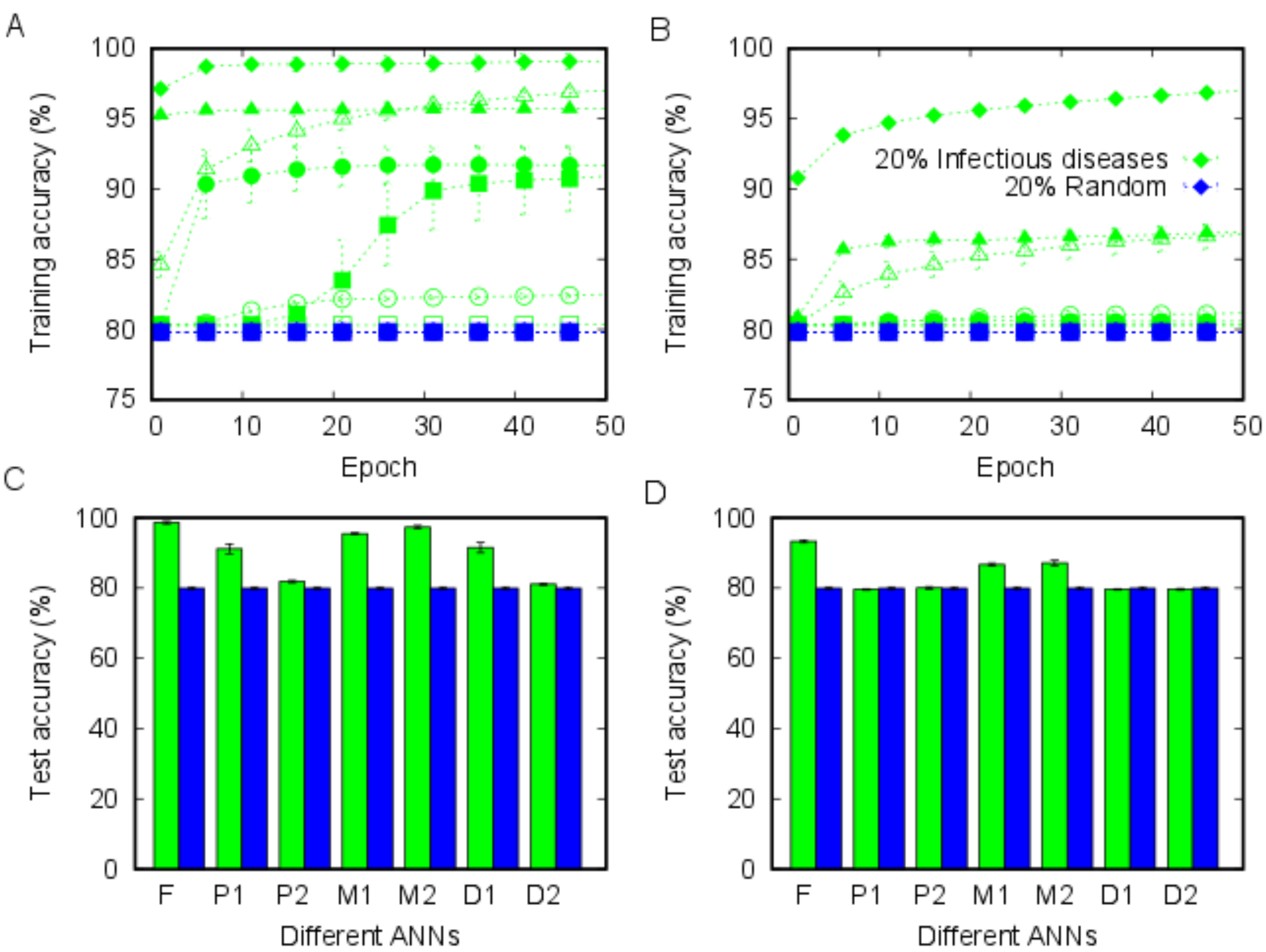

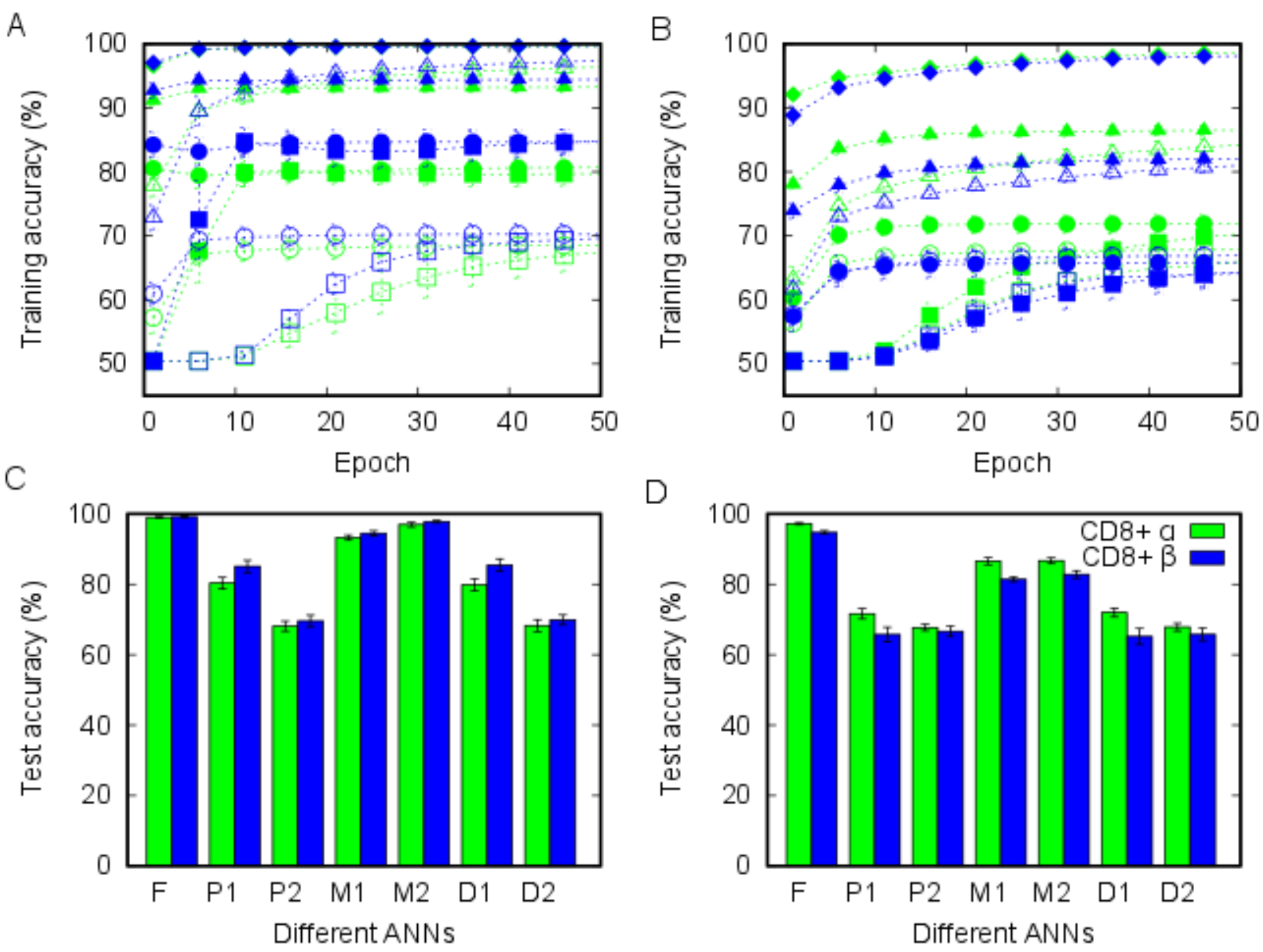